\documentstyle[eqsecnum,prb,aps,epsf]{revtex} 
\begin{document} \draft
\preprint{\today} \title{Boundary Condition Changing Operators in Conformal
Field Theory and Condensed Matter Physics
\thanks{For the proceedings of the conference ``Advanced
Quantum Field Theory'' (in memory of Claude Itzykson), La Londe les Maures,
France, September, 1996.}} \author{Ian
Affleck} \address{Department of Physics and Astronomy and Canadian
 Institute for Advanced
Research, \\ University of British Columbia, Vancouver, BC, 
Canada, V6T 1Z1} \maketitle \begin{abstract} Boundary condition changing
 operators
in conformal field theory describe various types of ``sudden switching''
problems in condensed matter physics such as the X-ray edge singularity. 
We review this subject and give two extensions of previous work.  A
general derivation of a connection between the  X-ray edge singularity, the
Anderson orthogonality catastrophe and finite-size scaling of energies is
given.  The formalism is also extended to include boundstates.  
\end{abstract}
\section{Introduction} Claude Itzykson has been an inspiring and supportive
colleague to me, ever since I had the good fortune to spend a year as his
neighbour at C.E.N. Saclay.  He is sorely missed.  

Conformal field theory with boundaries, a subject developed largely by John
Cardy,\cite{Cardy} has found applications to various quantum impurity problems
in condensed matter physics.  A particular branch of this subject involves
boundary condition changing operators.  They were used to obtain a new
understanding of a class of problems in condensed matter physics where a
local change in a Hamiltonian is made suddenly, for example by X-ray
absorption in a metal.  This article is mainly a review of work on this
subject by Andreas Ludwig and the author\cite{Affleck1} with a few new
insights presented.  In the next section we review boundary condition changing
operators in conformal field theory.  In section III we discuss their
application to the X-ray edge singularity and Anderson's orthogonality
catastrophe, giving some new results.  In the final section we discuss  X-ray
edge singularities in the Kondo problem.  Some of the new results were
obtained in collaboration with Alex Zagoskin.\cite{Zagoskin}
\section{Boundary Condition
Changing Operators in Conformal Field Theory}
Consider conformal field theory on the half-plane,\cite{Cardy} $z=r+i\tau$,
$r\geq 0$.  The requirement that the boundary of the half-plane, the real
axis, remain fixed reduces the full group of conformal transformations, $z\to
w(z)$, to those obeying $w(\tau)^*=w(\tau )$, an infinite subgroup.  We will
be interested in general conformally invariant boundary conditions on the
strip which are consistent with 
\begin{equation} T(\tau ,0)= \bar T (\tau ,0),\end{equation}
where $T$, ($\bar T$), is the holomorphic (anti-holomorphic) part of the
energy-momentum tensor.  This
corresponds to vanishing momentum density at the boundary, corresponding to
conservation of quantum mechanical probability density.  For a given bulk
conformal field theory, the set of conformally invariant boundary
conditions, $A$, $B$, ... can be determined.  

\begin{figure}
\epsfxsize=10 cm \centerline{\epsffile{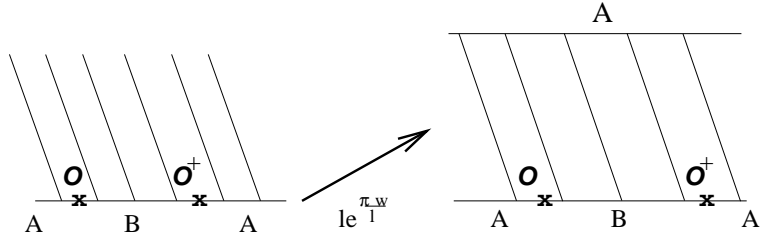}} 
\caption{Conformal transformation from the infinite half-plane to the strip.}
\label{fig:contran}
\end{figure}

In general the scaling dimensions of boundary operators can be related to the
finite size scaling of energy levels on a strip, as shown in Figure 
(\ref{fig:contran}).  Under
the conformal transformation $z=le^{\pi w/l}$ the correlation function of the
primary boundary condition changing boundary operator $O$ transforms as:
\begin{equation} {1\over (\tau_1-\tau_2)^{2x}}\to {1\over \left[{2l\over
\pi}\sinh {\pi \over 2l}(u_1-u_2)\right]^{2x}}
=\sum_n|<AA;0|O|AB;n>|^2e^{-[E^{AB}_n-E_0^{AA}]\Delta u}.
\label{main}\end{equation}
In the limit of large $\Delta u\equiv u_1-u_2$, we see that:
\begin{equation} 
{1\over \left[{2l\over
\pi}\sinh {\pi \over 2l}(u_1-u_2)\right]^{2x}}\to 
\left({\pi \over l}\right)^{2x}
e^{-\pi x\Delta u\over l}.\end{equation}
Note that $O$ changes the boundary condition from $A$ to $B$ on the lower
boundary.  Since the finite size energies, $E^{AB}_n$, depend on the
boundary conditions, $A$ and $B$, it follows that the dimensions of boundary
operators depend on the boundary conditions:
\begin{equation} x={l\over
\pi}[E^{AB}_n-E^{AA}_0],\label{energy}\end{equation} where $n$ labels the
lowest energy state with a non-zero matrix element in the sum in Eq.
(\ref{main}). Depending on the operator, $O$, $n$ may equal $0$,
corresponding to the groundstate with boundary conditions $AB$, or possibly
on excited state.  We also see from Eq. (\ref{main}) that this matrix element
scales with system size as: \begin{equation}|<AA;0|O|AB;n>|=\left({\pi\over
l}\right)^{x} \label{matem}\end{equation}

\section{application of boundary conformal field theory to the X-ray edge
singularity and Anderson's orthogonality catastrophe}
\begin{figure}
\epsfysize=5 cm \centerline{\epsffile{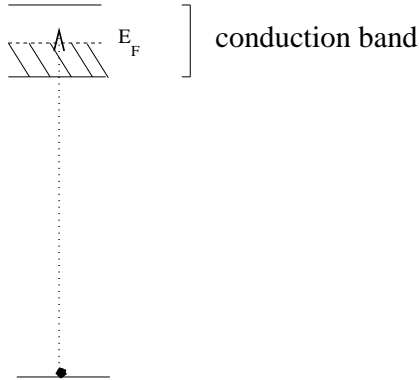}}
\caption{Core level and conduction band in a metal.}
\label{fig:core}
\end{figure}

\begin{figure}
\epsfysize=5 cm \centerline{\epsffile{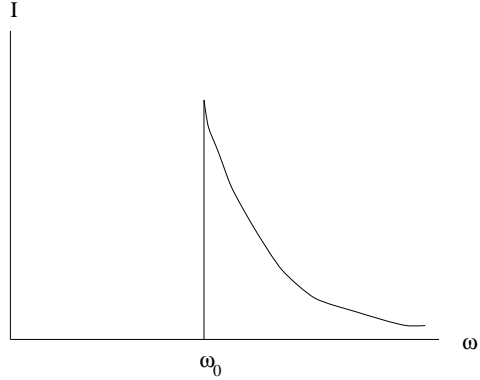}} 
\caption{Schematic X-ray absorption intensity versus frequency.}
\label{fig:thresh}
\end{figure}

Consider the electron energy levels in a metal. [See Figure 
(\ref{fig:core}).]
\noindent In addition to
the partially filled conduction band there are also core levels in which the
electron is tightly bound to a particular nucleus.  X-ray absorption can
either eject a core electron (photoemission) or raise it to the conduction
band, as indicated in Figure (\ref{fig:core}).
  The resulting core hole produces a local electrostatic potential
acting on the conduction electrons which affects the X-ray absorption
probability producing a singularity at the threshold, as indicated in Figure
(\ref{fig:thresh}).  A very simple model was discussed by Nozi\`eres and De
Dominicis.\cite{Nozieres}
 \begin{equation} H=\sum_{\vec k}\epsilon_{\vec k}c_{\vec
k}^\dagger c_{\vec k} +b^\dagger b\sum_{\vec k,\vec k'}V_{\vec k,\vec
k'}c_{\vec k}^\dagger c_{\vec k'}-E_0b^\dagger b.\end{equation}
Here $\epsilon_{\vec k}$ is the dispersion relation of the conduction band
electrons, annihilated by $c_{\vec k}$.  $b$ annihilates the core electron. 
$V$ is the local potential. (Electron spin is ignored.)  An essential
observation\cite{Nozieres} is that: \begin{equation} [H,bb^\dagger
]=0.\end{equation} The Hilbert space consists of two sectors.  In one
$bb^\dagger =0$; the core level is filled and there is no potential.
In the other  $bb^\dagger =1$; the core level is empty and
consequently the potential $V$ is exerted on the conduction electrons.  In
the simpler case of X-ray photoemission, the absorption intensity is:
\begin{equation} I(\omega )\propto \int dte^{i\omega t}<0|b^\dagger
(t)b(0)|0>\propto (\omega -\omega_0)^{-\alpha}.\end{equation}Inserting the $b$
operator switches on the potential, $V$, at time 0.  It is later switched off
at time $t$.  At low energies, turning on  $V$ is equivalent to changing the
boundary conditions.  Thus $b$ is a boundary condition changing operator.  

\begin{figure}
\epsfxsize=10 cm \centerline{\epsffile{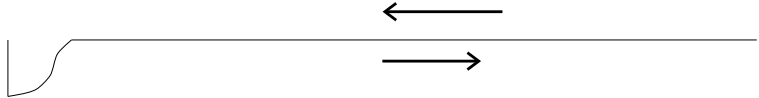}} 
\caption{One-dimensional problem with potential at the origin.}
\label{fig:1D}
\end{figure}

We will assume that both $\epsilon_{\vec k}$ and $V_{\vec k,\vec k'}$
are spherically symmetric and consider only s-wave scattering. (It is
straightforward to generalize the method to other cases.) 
Linearizing the dispersion relation around the Fermi surface, we obtain a
low energy theory consisting of (1+1) dimensional Dirac fermions defined on
the half-line $r>0$, with the potential $V$ at the origin.  See Figure 
 (\ref{fig:1D}).
We integrate out all degrees of freedom except for electrons
with energies very close to the Fermi energy, $E_F$.  Thus the fields obey
an effective boundary condition at the origin:
\begin{equation} \psi_R(0)=e^{2i\delta
(E_F)}\psi_L(0),\label{bc}\end{equation}
where $\delta (E_F)$ is the s-wave phase shift at the Fermi energy.

The state $b|0>$ has the core hole present but the conduction electrons
filling the V=0 Fermi sea.  In the infinite volume limit, this state is
orthogonal to the groundstate with the core hole present, in which  the
electrons in the Fermi sea are rearranged by the potential.  This is known
as the ``Anderson orthogonality catastrophe''.  It is this rearrangement of
the Fermi sea, produced by V, which leads to the X-ray edge singularity.  

On a finite strip (corresponding to s-wave electrons inside a finite
sphere):
\begin{equation} <AB;0|b|AA;0>\neq 0.\end{equation}
Here $|AA;0>$ is the unperturbed filled Fermi sea and $|AB;0>$ is the
state with the core electron absent and the filled Fermi sea perturbed by the
presence of the core hole potential, V.  From the conformal transformation
of Eq. (\ref{main}) we see that the X-ray exponent, the
Anderson orthogonality exponent and the $1/l$ term in the groundstate energy
difference are all given by the same number, $x$:
\begin{eqnarray}<b^{\dagger} (t)b (0)>&\approx& {1\over t^{2x}}\ \
[\hbox{X-ray exponent}\ \alpha =1-2x]\nonumber \\
|<AB;0|b|AA;0>|&=& \left({\pi\over l}\right)^x \ \ [\hbox{orthogonality
exponent}]\nonumber \\
E^{AB}_0-E^{AA}_0 &=&{\pi x\over l} \ \ [\hbox{finite size
energy}].\label{gx}\end{eqnarray}
The exponent, $x$, is most easily calculated from the finite size energy. 
It is instructive to calculate the complete expression for this energy
difference, without making any low energy approximations, in order to ensure
that the $1/l$ term indeed only depends on the phase shift at the Fermi
surface, $\delta (E_F)$, that is on the boundary condition in the low
energy effective theory.  For large $l$, \begin{equation} E^V_0-E^0_0\approx
l\int_0^{k_F}{dk\over \pi}[\epsilon (\tilde k)-\epsilon (k)],\end{equation}
where $k_F$ is the Fermi wave-vector and the phase-shifted wave vector,
$\tilde k$, is defined by the {\it self-consistent} equation:
\begin{equation}\tilde k \equiv k-{\delta(\tilde k)\over l}.\end{equation} 
Taylor expanding out to terms of $O(1/l)$, \begin{eqnarray} \Delta E&\approx
&-\int_0^{k_F}{dk\over \pi} \left[-\epsilon '(k)\delta (k)+{1\over
2l}\epsilon ''(k)\delta (k)^2+{1\over l} \epsilon '(k)\delta '(k)\delta
(k)\right]\nonumber \\ &=&-{1\over \pi}\int_0^{E_F}d\epsilon \delta (\epsilon
)+v_F{\pi \over l}{1\over 2}\left[{\delta (E_F)\over \pi}
\right]^2.\label{xdelta}\end{eqnarray} Here $v_F=\epsilon '(k_F)$ is the Fermi
velocity and we have integrated by parts to obtain the last line, using the
fact that $\epsilon '(0)=0$ which follows from analyticity of the dispersion
relation.  The first term in $\Delta E$ is O(1) and depends on the phase
shift thoughout the conduction band.  This formula is known as Fumi's
theorem.  On the other hand, the second term, of $O(1/l)$, only depends on
the phase shift right at the Fermi surface.  Thus this term (although not the
first one) is given correctly by the low energy theory which only keeps
states near the Fermi surface.  It is determined by the boundary condition of
Eq. (\ref{bc}).  Taking into account that we have set $v_F=1$ in our previous
equations, we see that the scaling dimension of the boundary condition
changing operator, $b$ is given by:
\begin{equation} x={1\over 2}\left[{\delta (E_F)\over \pi}
\right]^2.\label{x}\end{equation}

\begin{figure}
\epsfxsize= 10 cm \centerline{\epsffile{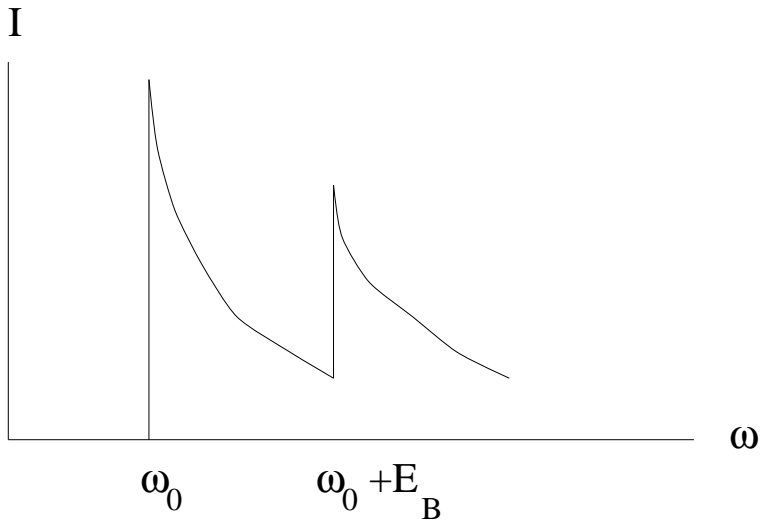}} 
\caption{Schematic X-ray absorption intensity versus frequency when the
core potential produces a boundstate.} \label{fig:bs}
\end{figure}

Now suppose that the potential V is attractive and produces a boundstate. 
In this case, the X-ray absorption intensity has 2 thresholds: a lower energy
one corresponding to intermediate states in which the boundstate is occupied
and a higher energy one corresponding to intermediate states in which it is
empty.\cite{Combescot}  [See Figure  (\ref{fig:bs}).]
These are separated by the binding
energy of the boundstate, $E_B$.  A crucial observation is that  the
boundstate wavefunction and energy decay exponentially and hence don't
contribute to the $O(1/l)$ terms in the energies.  Hence the exponent at the
first threshold is given by precisely the previous formulas, Eqs.
(\ref{gx}, \ref{x}). To obtain the exponent at the second
threshold, where the boundstate is empty, we just need to calculate the
$1/l$ term in the energy.  The calculation is precisely the same as before
except that now there is one extra electron in the continuum, since it is not
in the  boundstate. The corresponding boundary condition changing operator in
this case has no matrix element to the groundstate, $|AB;0>$ so the leading
term in Eq. (\ref{main}) comes from the excited state with one extra electron
at the Fermi surface.  The next available state at the Fermi surface has
wave-vector $(\pi /2-\delta (E_F))/l$, so:
\begin{eqnarray} \Delta E &\to &\Delta E+\epsilon \left[ k_F+{\pi /2-\delta
(E_F)\over l}\right] \nonumber \\ &=& \Delta E + E_F+v_F{\pi \over
l}\left({1\over 2}-{\delta (E_F)\over \pi}\right)\nonumber \\
&=&-{1\over \pi}\int_0^{E_F}d\epsilon \delta (\epsilon )+E_F+v_F{\pi \over
l}{1\over 2}\left[{\delta (E_F)\over \pi}-1
\right]^2.
\end{eqnarray}    Hence the
scaling dimensions for the filled and empty boundstate thresholds, $x_f$ and
$x_e$ are: \begin{equation} x_f={1\over 2}\left[{\delta (E_F)\over \pi }
\right]^2,\ \ 
x_e={1\over 2}\left[{\delta (E_F)\over \pi} -1\right]^2.\end{equation}

These results for the X-ray exponent were all obtained earlier by Nozi\`eres
and De Dominicis\cite{Nozieres} and Combescot and
Nozi\`eres\cite{Combescot}.  An early calculation using bosonization was
given by Schotte and Schotte.\cite{Schotte}  The result for the orthogonality
exponent (ignoring boundstates) was obtained by Anderson.\cite{Anderson} 
Nonetheless, the present approach, based on conformal transformation and
calculation of the energy difference, has several advantages.  Once the
relationship to the energy difference is obtained, pages of complicated
calculations reduce to a few lines.  In particular, using previous
approaches, it is only straightforward to obtain the relationship between
exponents and phase shift in the Born approximation, where the phaseshift is
proportional to the (weak) potential, $V$.  The simple derivation given above
[Eq. (\ref{xdelta})] establishes, via the connection with  groundstate
energies, that it is indeed precisely the phaseshift at the Fermi surface,
$\delta (E_F)$, which determines the exponents, even when the potential is
not weak.  Furthermore, our new approach easily permits numerous
generalizations to less trivial problems.  For example, it can be used to
calculate X-ray and orthogonality exponents in an interacting one-dimensional
electron gas (Luttinger liquid)\cite{Affleck1}.  As long as the bulk
interactions leave the electrons gapless, the low energy theory is a
conformal field theory (free massless boson).  While backscattering from the
core hole potential leads to highly non-trivial renormalization, in the end
the low-energy physics is governed by a simple conformally invariant boundary
condition, for which the value of the scaling dimension, $x$, can
be determined from the finite-size energies, 
Eq. (\ref{energy}).\cite{Wong}  Recent density matrix
renormalization group work\cite{Qin} confirms the prediction for the
exponent, $x$, from both the finite-size energy and the matrix element, in
Eq. (\ref{gx}).  On the other hand, a direct numerical simulation of the
Green's function in Eq. (\ref{gx}) is a much more formidable task.  We review
another non-trivial application of this approach in the next section.

\section{the kondo problem and the X-ray edge singularity}
Now we endow both conduction electrons and core electrons with a spin index,
$\alpha$, $c_{\vec k \alpha}$, $b_{\alpha}$.  As well as potential scattering
there could also be a spin-spin interaction between core and conduction
electons:
\begin{equation} H \to H +b{\vec \sigma \over 2}b^\dagger \cdot \sum_{\vec
k, \vec k'}J_{\vec k, \vec k'}c_{\vec k}{\vec \sigma \over 2}c_{\vec
k'}.\label{H_K}\end{equation}
(A sum over spin indices is implied.)
  As before, $[H,bb^\dagger ]=0$.  When the core
hole is present, it acts like an s=1/2 impurity spin:
\begin{equation} \vec S_{\hbox{imp}}\equiv b{\vec \sigma \over
2}b^\dagger.\end{equation}
When it is absent the core level is spinless.  Thus the X-ray absorption
turns on a Kondo interaction, $J_{\vec k, \vec k'}$.  This problem appears
considerably more difficult than the potential scattering case, discussed in
the previous section, because the Kondo Hamiltonian contains non-trivial
physics even without sudden switching.  Assuming J is antiferromagnetic
($>0$), it renormalizes to large values in the infrared.  However, it has
been shown that the infrared stable fixed point corresponds to a conformally
invariant boundary condition.\cite{Affleck2}  From this solution of the Kondo
problem we can determine X-ray edge and orthogonality exponents.

The conformally invariant boundary condition giving the infrared fixed point
can be specified in terms of the finite size spectrum.\cite{Affleck2} We
begin with free boundary conditions on both sides of a finite strip.  This
leads to a trivial spectrum.  It is convenient to express this in terms of a
(trivial) conformal embedding.  Beginning with two Dirac fermions (one for
each spin) we bosonize and introduce charge and spin bosons.  The finite
size spectrum can be expressed in terms of $SU(2)_1$ Kac-Moody conformal
towers with highest weight states of spin $j=0,1/2$.  The reason for two
SU(2)'s is that the underlying theory of 2 free Dirac fermions is equivalent
to 4 free Majorana fermions and hence has SO(4) symmetry.  Expressed in
terms of $SU(2)_1$ characters, the spectrum is:
\begin{equation}Z_{\hbox{free,free}}=\chi_0\chi_0+\chi_{1/2}\chi_{1/2},
\end{equation}
where the first and second factor come from the spin and charge sector
respectively.  The infrared Kondo fixed point corresponds to fusion with the
$j=1/2$ primary field in the spin sector.  This fusion process is a
conformal field theory representation of screening of the (s=1/2) impurity
by the conduction electrons.  The resulting spectrum is:
\begin{equation} Z_{\hbox{Kondo,
free}}=\chi_{1/2}\chi_0+\chi_0\chi_{1/2}.\end{equation} 
The energies of highest weight states, or scaling dimensions of primary
operators, for spin $j$ and $SU(2)_k$ Kac-Moody algebra are given
by:\cite{Knizhnik} \begin{equation} x_{j,k}={j(j+1)\over 2+k}. 
\label{dimkj}\end{equation}
The core hole operator, $b_\alpha$, takes the groundstate into the state
with spin 1/2 and charge 0.  (Only the charge of conduction electrons is
relevant.)  It has scaling dimension 
\begin{equation} x_{1/2,1}={1\over 4}.\end{equation}

So far we have only considered photoemission in which the core electron is
ejected from the metal.  Now we consider the case where instead the core
electron is excited into the conduction band.  Ignoring the eventual decay
back into the core level (which occurs incoherently) the X-ray aborption
intensity in this case is given by:
\begin{equation} I(\omega )\propto \int dt e^{i\omega t}<b^{\dagger \alpha}
(t) c_\delta (t) c^{\dagger \beta}(0)b_\alpha (0)>.\end{equation}
In principle (not worrying about possible selection rules in the X-ray
absorption process), the core hole and extra conduction band electron can be
in a singlet or triplet state.  Different exponents are obtained for these
two cases.  We must calculate the dimension of the boundary condition
changing operators $c^\dagger \vec \sigma b$ and $c^\dagger b$. 
These change the groundstate into a state with Kondo boundary conditions and
quantum numbers $Q=1$, $j=1$ and $Q=1$, $j=0$ respectively. (We only count
the charge of the conduction electrons.)  The exponent, $x$, is obtained
from the lowest energy state with the right quantum numbers in the
$Z_{\hbox{Kondo, free}}$ partition function.  $Q=1$, $j=0$ corresponds to a
primary field, of dimension 
\begin{equation} x_0={1\over 4},\end{equation}
 from Eq. (\ref{dimkj}), since  $Q=1$ corresponds to the $j=1/2$ primary
of a second $SU(2)_1$ KM algebra associated with charge.  On the
other hand, their is no primary with j=1 for $SU(2)_1$.  This state is a
descendent in the $Q=1$, $j=0$ product of conformal towers, with dimension:
 \begin{equation} x_1={1\over 4} + 1={5\over
4}.\end{equation} 
(It is a Mac-Moody descendent but a Virasoro primary, so Eq. (\ref{main})
still applies.)  The corresponding X-ray edge exponents are obtained from
the usual formula, $\alpha = 1-2x$.  

This calculation can be generalized to the k-channel Kondo problem in which
the conduction electrons carry an additional flavour index, $c_{\alpha i}$,
$i=1,2,3,... k$.  The Kondo interaction with the core spin (which doesn't
carry a flavour index) is as in Eq. (\ref{H_K}) with a diagonal sum over the
flavour of the conduction electrons.  The infrared fixed point for the
multichannel Kondo problem has also been indentified as a conformally
invariant boundary condition.\cite{Affleck2}
  A key step in this identification is a
conformal embedding of 2k Dirac fermions.  One first uses non-abelian
bosonization to represent the 2k fermions by a charge boson and a $SU(2k)_1$
Wess-Zumino-Witten model.\cite{Witten,Knizhnik}  Then one uses the conformal
embedding: \begin{equation} SU(2k)_1\to SU(2)_k\times SU(k)_2,\end{equation}
which was worked out by Altsh\"uler, Bauer and Itzkson.\cite{Altshuler}
[The $SU(k)_2\times U(1)$ theory is actually equivalent\cite{Affleck3} to
$SP(k)_1$, but we won't use this result here.]  The finite size
spectrum with Kondo-free boundary conditions is again obtained by fusion with
the j=1/2 conformal tower in the $SU(2)_k$ theory.  The photoemission
exponent is given by the dimension of $b_\alpha$,  with j=1/2, Q=0. From Eq.
(\ref{dimkj}) this is: \begin{equation} x={3/4\over 2+k}.\end{equation}
The operators giving the X-ray exponents for the case where the core
electron goes into the conduction band, $c^{i\dagger}\vec \sigma b$ and
$c^ib$  have j=1 or 0 respectively, Q=0 and
transform under the fundamental representation of $SU(k)$.  From the fusion
transformation on the free fermion spectrum, expressed using the conformal
embedding, it can be checked that the corresponding primaries occur in the
$Z_{\hbox{Kondo, free}}$ spectrum.  The scaling dimensions are:
\begin{equation} x={1\over 4k}+{k^2-1\over 2k(2+k)}+{j(j+1)\over
2+k},\ \  (j=0 \ \hbox{or}\  1). \end{equation} 

\acknowledgements I would like to thank my collaborators in this work,
 Andreas Ludwig and Alex
Zagoskin.  I also thank Jeff Young for interesting me in edge singularities in
systems with boundstates.  This research was supported in part by NSERC of
Canada.  


\begin{references} 
\bibitem{Cardy} J.L. Cardy, Nuc. Phys. {\bf B324}, 581 (1989).
\bibitem{Affleck1} I. Affleck and A.W.W. Ludwig, J. Phys. {\bf A27}, 5375
(1994).
 \bibitem{Zagoskin} A. Zagoskin and I. Affleck, in preparation.
\bibitem{Nozieres} P. Nozi\`eres and C.T. De Dominicis, Phys. Rev. {\bf
178}, 1097 (1969).
\bibitem{Combescot} M. Combescot and P. Nozi\`eres, J. de Physiques {\bf
32}, 913 (1971).
\bibitem{Schotte}K.D. Schotte and U. Schotte, Phys. Rev. {\bf 182}, 479
(1969).
\bibitem{Anderson}P.W. Anderson, Phys. Rev. {\bf 164}, 352 (1967).
\bibitem{Wong} E. Wong and I. Affleck, Nucl. Phys. {\bf B417}, 403 (1994).
\bibitem{Qin} S. Qin, M. Fabrizio and L. Yu, preprint: cond-mat/9608024.
\bibitem{Affleck2} For a review see I. Affleck, Acta Physica Polonica {\bf
26}, 1869 (1995); cond-mat/9512099.
\bibitem{Knizhnik}V. Knizhnik and A. Zamolodchikov, Nucl. Phys. {\bf B247},
83 (1984).
\bibitem{Witten} E. Witten, Comm. Math. Phys. {\bf 92}, 455 (1984).
 \bibitem{Altshuler} D. Altsch\"uler, M. Bauer and C. Itzykson, Comm.
Math. Phys. {\bf 132}, 349 (1990).
\bibitem{Affleck3} I. Affleck, A.W.W. Ludwig, D.L. Cox  and H.-B. Pang,
Phys. Rev. {\bf B45}, 7918 (1992).
 \end{references}
 \end{document}